# Binding Sites, Vibrations and Spin-Lattice Relaxation Times in Europium(II)-based Metallofullerene Spin Qubits

Ziqi Hu[1,2], Aman Ullah[1], Helena Prima-Garcia[1], Sang-Hyun Chin[1], Yuanyuan Wang[2], Juan Aragó[1], Zujin Shi[2*], Alejandro Gaita-Ariño[1*] and Eugenio Coronado[1*]

[1] Instituto de Ciencia Molecular, Universidad de Valencia, C/Catedrático José Beltrán 2, 46980 Paterna, Spain.

[2] National Laboratory for Molecular Sciences, State Key Laboratory of Rare Earth Materials Chemistry and Applications, College of Chemistry and Molecular Engineering, Peking University, Beijing 100871, People's Republic of China.

**ABSTRACT:** To design molecular spin qubits with enhanced quantum coherence, a control of the coupling between the local vibrations and the spin states is crucial, which could be realized in principle by engineering molecular structures via coordination chemistry. To this end, understanding the underlying structural factors that govern the spin relaxation is a central topic. Here, we report the investigation of the spin dynamics in a series of chemically-designed europium(II)-based endohedral metallofullerenes (EMFs). By introducing a unique structural difference, i.e. metal-cage binding site, while keeping other molecular parameters constant between different complexes, these manifest the key role of the three low energy metal-based vibrations in mediating the spin-lattice relaxation times ($T_1$). The temperature dependence of $T_1$ can thus be normalized by the frequencies of these low energy vibrations to show an unprecedentedly universal behavior for EMFs in frozen $CS_2$ solution. Our theoretical analysis indicates that this structural difference determines not only the vibrational rigidity but also spin-vibration coupling in these EMF-based qubit candidates.

## INTRODUCTION

Electronic spins promise a yet untapped potential as nanoscale memories, both as classical bits[1] and as quantum bits (spin qubits)[2]. The role of chemistry to overcome this interdisciplinary challenge is to offer design strategies from a bottom-up approach[3]. This requires both chemical control to come up with synthetic routes towards the desired structures and physical insights to define the goals. In the case of spin qubits, one of the main challenges on our path is quantum decoherence[4]: the loss of quantum information that in chemical terms can be seen simply as relaxation, which is often characterized by pulsed electron paramagnetic resonance (EPR) and allows us to obtain information on the spin-lattice ($T_1$) and spin-spin ($T_2$, measured as phase memory time $T_m$) relaxation times. Chemical strategies to extend $T_2$ include the design of molecular architectures that are free from nuclear spins[5], dilution within a diamagnetic matrix[6], or choosing a crystal field Hamiltonian which allows for "atomic clock transitions", that in turn protect the spin states from magnetic noise[7]. However, a necessary condition for a long $T_2$ is a long $T_1$; in simple terms, preserving quantum information is only possible if the classical memory is also preserved. In practice, this means designing molecules where the spin states are protected from vibrations, and a great deal of effort has been invested in this[8]. A major challenge in these investigations is to find a chemical platform in which controlled changes can be introduced with the aim of varying the relevant parameters at will.

Magnetic fullerenes, which have been characterized as molecular spin qubits[9], are exceptional model systems in this context due to their beautiful chemical and structural simplicity. Among these, endohedral metallofullerenes (EMFs) offer the possibility of including entrapped magnetic ions,[10] which occupy specific positions in the carbon cage and have distinct coordination environments thanks to the structural diversity offered by the cages and the availability of binding sites. The nature and geometry of the binding sites are chosen from a small set since these nanostructures are entirely composed of carbon atoms forming hexagons and pentagons. Moreover, this extraordinary simplicity even extends to the vibrational modes: $M^{III}@C_{82}$ (M = Y, La, Ce, Gd) has been reported to exhibit a vibrational spectrum where the three relative motions between metal ion and carbon cage show the lowest frequencies compared with all other cage distortions[11]. To probe how metal-cage binding influences the magnetic properties of EMFs, intermolecular interactions should be effectively quenched; otherwise, the different shapes of the cages may play a role by governing their packing. Thus, isolating EMF molecules from each other using a fairly rigid and simple solvent like $CS_2$, which is also very poor in nuclear spins, would make this system close to a gas-phase ideal situation. This lays out a perfect scenario for a systematic study.

In the present work we investigate the role of vibrations in the relationship between molecular structure and thermally-activated spin relaxation. Towards this end, divalent europium ion with a $4f^7$ electronic configuration is an excellent candidate due to (*i*) its well-isolated ground spin multiplet $S = 7/2$ and (*ii*) closed-shell structure of its hosting carbon cage, leading to the simplest possible spin Hamiltonian. Here, we report four novel spin qubits based on divalent monoeuropium EMFs using a combination of experimental and theoretical characterization techniques, including density functional theory (DFT) to model molecular vibrations, pulsed EPR to study the spin dynamics, and complete active space self-consistent field theory (CASSCF) to model the evolution of the spin energy levels along with vibrational distortions.

## RESULTS AND DISCUSSION

**Structural and vibrational analysis.** A total of 10 $Eu@C_{2n}$ (2n = 74-90) were synthesized and purified according to the conventional procedure[12] and characterized by laser desorption time-of-flight (LD-TOF) mass spectroscopy, vis-near-infrared (Vis-NIR) absorption and photoluminescence (see experimental section and Supporting Information). As is typical for EMFs, while the structures consist in pentagons and hexagons, the overall geometry presents in most cases a low symmetry, and the metal ion —having ample room inside the carbon cage— is attached to a wall; in particular, it is bonded to a specific site that varies from case to case and depends non-trivially on the structure of the carbon cage[12-13]. Accordingly, four typical binding sites, namely of types: "hexagon", "acephenalene", "pyracylene" and "fused pentagons", can be classified for all $Eu@C_{2n}$ molecules (**Figure 1a**). Note that in acephenalene type, the metal ion is closer to the pentagon-hexagon-hexagon junction[13c]. As representative molecules we chose $Eu@C_{84}$-$C_2$(13) (**1**), $Eu@C_{82}$-$C_s$(6) (**2**), $Eu@C_{74}$-$D_{3h}$(5) (**3**) and $Eu@C_{76}$-$C_{2v}$(19138) (**4**), respectively. However, this study is focused on **1**, **2** and **3** since **4** was not obtained due to its low yield. In the spin dynamic studies discussed later, $Eu@C_{80}$-$C_{2v}$ (**3'**) is also included, which possesses the same pyracylene binding site as for **3**. DFT optimized structures of $Eu@C_{2n}$ molecules are depicted in **Figure 1a** and **Figures S2.1-2.4**.

Let us for the moment set aside the difference between these four classes of EMFs and focus on what they have in common. DFT calculations on all $Eu@C_{2n}$ molecules showed that the three vibrational modes with the lowest frequencies consistently correspond to displacements of the metal ion relative to the carbon cage along $x$-, $y$- and $z$- directions (**Figure 1a-b**). Since we are taking the $z$ axis as the one defined along the contact between the surface of the carbon cage and the metal ion, we will refer to the $z$ distortion as "longitudinal" and to the $x$ and $y$ distortions as "lateral". Lateral distortions are the softest ($v < 70$ cm$^{-1}$) and purely involve a movement of the metal ion along the inner wall of the carbon cage. In contrast, longitudinal distortions strain the interaction between the metal and the cage and, as a result, they are slightly stiffer ($v \approx 125$ cm$^{-1}$). These vibrational modes also involve a slight distortion of the carbon cage, in the sense that the binding site is "tugged down" by the metal ion as it moves away from the surface. All the other vibrational modes are purely cage distortions (**Figure S2.6**) with higher frequencies ($v > 210$ cm$^{-1}$, **Table S2-6**). In all of these, the relative position of the metal ion and its nearest neighbors is kept approximately constant, as evidenced by the reduced masses of these modes being very close to 12 u.m.a.q.. Thus, our working hypothesis is that only lateral and longitudinal vibrational modes are significantly coupled with the energies of the spin states in Eu$^{2+}$, not only because these are the only ones that significantly alter the coordination environment of Eu$^{2+}$, but also because they are the ones that can be thermally populated at relatively low temperatures.

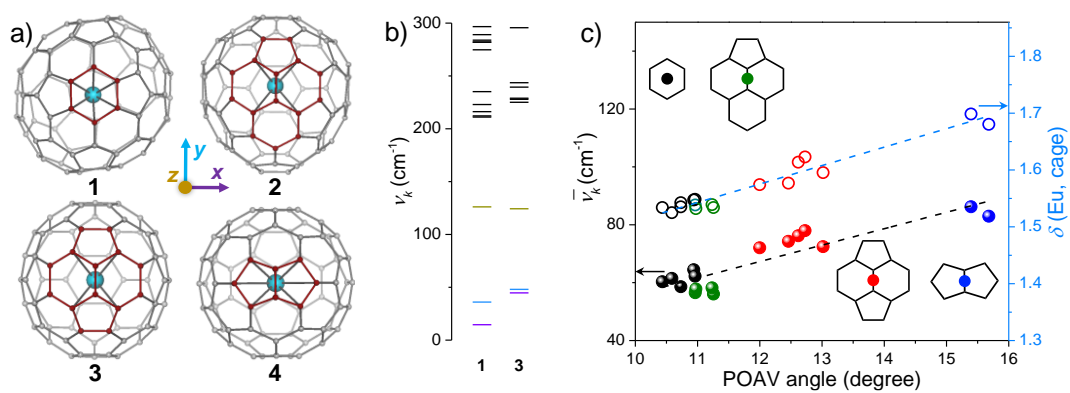

**Figure 1. a)** Top views of the optimized molecular structures of **1**, **2**, **3** and **4**. Color code: light blue, Eu; grey, C. The binding sites are highlighted in red. The inset indicates that the first three vibrational modes feature metal-to-cage motions along $x$-, $y$- (lateral) and $z$-directions (longitudinal). **c)** Calculated vibrational energy levels for **1** and **3** below 300 cm$^{-1}$. Black indicate cage-dependent vibrations. **c)** Correlations between averaged $p$-orbital axis vector (POAV) of binding sites with averaged frequencies $\overline{v_k}$ ($k$ = 1-3) of the three metal-based vibrations (left axis, solid symbols), and with metal-cage delocalization indices $\delta$(Eu, cage) (right axis, open symbols). The linear fits are shown as dashed lines. POAV angle of hexagon is determined as the average value of the six carbon atoms. The other three structures adopt the average of the two nearest carbon atoms. $\delta$(Eu, cage) is determined as the sum of all $\delta$(Eu, C) values. The labelling is according to the four binding sites as hexagon (black), acephenalene (green), pyracylene (red) and fused pentagons (blue). The insets show their schematic diagrams.

To quantify the chemical and structural differences between the four classes of binding sites, we employed the delocalization index $\delta$(Eu,cage), which parameterizes the number of the shared electron pairs between the Eu$^{2+}$ ion and the carbon cage[14], and the $p$-orbital axis vector (POAV) pyramidalization angle, which is a purely structural parameter that represents the curvature of a conjugated system, accounting for the deviation extent of a carbon atom from ideally planar $sp^2$ hybridization[15]. It is interesting to plot these two parameters against each other and to use them as backdrop to understand the evolution of the averaged vibrational frequency $\overline{v_k} = (v_1 + v_2 + v_3)/3$ (**Figure 1c**). The parallel evolution of the three parameters evidences the direct relation between structure, bond strength and vibrational stiffness. It is also easy to conclude in this analysis that **1** and **2** are very similar to each other, whereas **3** and **4** present progressively more curved binding sites, stronger bonds and stiffer vibrations. The key here to classify the binding structures is the pentagon in the coordination environment, as it is at the origin of curvature in spherical carbon architectures. Indeed, the maximum pentagon separation rule governs the structural stability of EMFs and the local motifs, with the curvature induced by neighboring pentagons, preferentially accept electrons from the internal metal[16]. Thus, "fused pentagons" and "pyracylene" binding sites lead to strong metal-cage interaction and concomitantly rigid vibrations, whereas flatter motifs such as "hexagon" and "acephenalene" result in a weaker bonding of the carbon cage to the metal ion, enhancing its mobility.

**EPR studies.** $Eu^{2+}$ is isoelectronic to $Gd^{3+}$, meaning it has a $4f^7$ electronic configuration, with an $^8S_{7/2}$ ground term. The degeneracy is broken by mixing with excited states, which can be characterized by a zero-field splitting (ZFS):

$$\hat{H}_S = D[\hat{S}_z^2 - \frac{1}{3}S(S+1)] + E(\hat{S}_x^2 - \hat{S}_y^2) + g_{iso}\mu_B \boldsymbol{B}\boldsymbol{\hat{S}}, \qquad (1)$$

where the first two terms are the second order axial and rhombic ZFS of the spin multiplet, $D$ and $E$, and the last term represents the Zeeman effect[17].

We employed X-band continuous-wave (CW) EPR spectroscopy in frozen $CS_2$ solution and in the magnetic field range of 0-12000 G to estimate $D$ and $E$ in **1**, **2** and **3**. The fits allow reasonable reproductions of the powder spectra (**Figure 2**). These results are confirmed by pulsed EPR measurements, employing the standard Hahn echo sequence ($\pi/2$-$\tau$-$\pi$-$\tau$-echo) to record the echo-detected field-sweep (EDFS) spectra. The combination of both techniques allows us to identify the peaks and fields at which to perform further spin dynamics investigations. The calculated spin energy level schemes are similar in **1** and **2**, as a result from their similar values of $D \approx 0.28$ cm$^{-1}$, whereas for **3** we recover a much smaller value $D = 0.13$ cm$^{-1}$. Although they are larger than those of some Gd(III)-based compounds, such as GdW$_{30}$ ($D = 0.043$ cm$^{-1}$)[17b] and Gd$_2$@C$_{79}$N ($D = 0.033$ cm$^{-1}$)[18], comparable $D$ values were determined in derivatized Gd@C$_{82}$(morpholine)$_n$ with n = 5, 7, 9 and $D$ values ranging from 0.22 cm$^{-1}$ to 0.31 cm$^{-1}$ [19].

We proceeded to study relaxation dynamics via pulsed EPR (see **Figure 3**). Let us start by describing the evolution of $T_1$ as a function of the studied molecule, the chosen peak within that molecule, and the temperature. One can immediately appreciate that within each system there is no substantial dependence on the studied peak, meaning transitions involving different spin energy levels relax at similar rates. In contrast, there is a stark temperature dependence, and also a marked difference between what are apparently two kinds of systems: in one category, **1** and **2**, with softer binding sites, and in a second category **3** and **3'**, with more rigid pyracylene binding site. Among the different models that we employed (**Figure S3.14**), the best fits are obtained in **Figure 3a** assuming a Raman process: $T_1 = C^{-1}T^{-n}$, where $n = 3$ for both categories.

For the soft binding sites, we obtain $C = 9.1(3) \times 10^{-4}$ $\mu s^{-1}K^{-3}$ and the fit is valid for all the studied temperature range, whereas for the rigid binding site $C = 1.5(1) \times 10^{-4}$ $\mu s^{-1}K^{-3}$ but the fit is only acceptable down to 6 K. This Raman-type relaxation has already been observed in nearly isotropic systems, namely a 4f single-ion qubit[20] and a spin-1/2 system[8g]. A remarkable insight can be extracted when the temperatures of each experiment are normalized by the averaged reciprocal vibrational frequencies of the studied molecule (**Figure 3a**, inset). This normalization gives rise to a dimensionless number that qualitatively informs us about the expected population of vibrational states. This practically results in an overlap of all the experimental points that correspond to the Raman mechanism, manifesting that the temperature dependence of $T_1$ is primarily related to the vibrational rigidity. It emerges from this analysis that the low energy vibrations are indeed important in mediating spin-lattice relaxation, which is in line with the observations for vanadium(IV)-based qubits[21]. Further, the frequency normalization approach may also be extended to this system, where the key low energy vibrations are well determined for analogous structures[22].

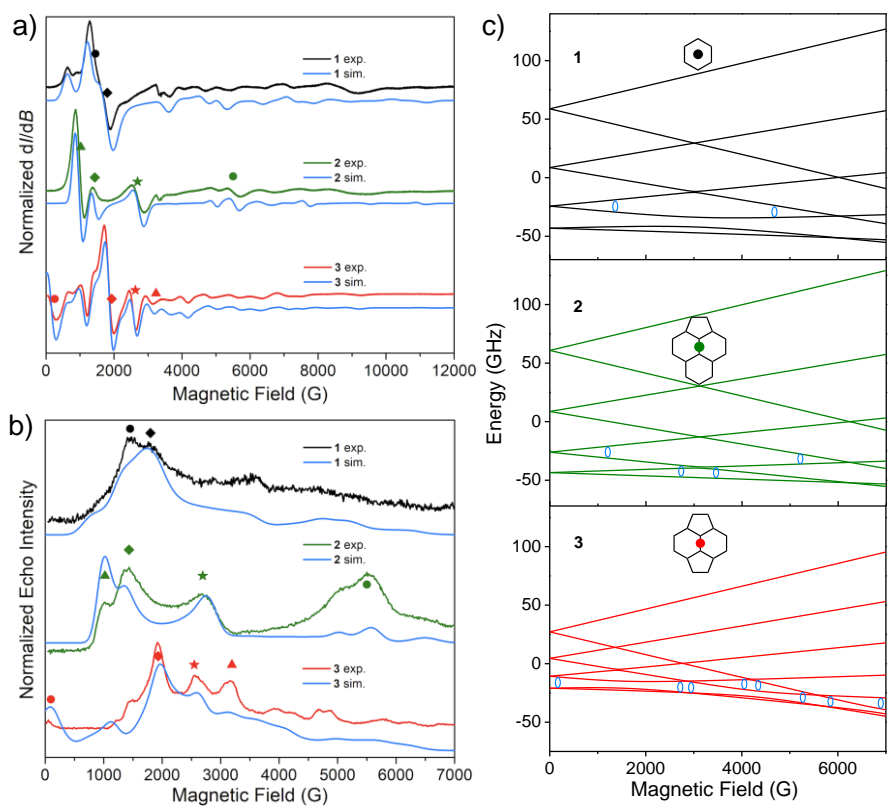

**Figure 2. a)** X-band CW-EPR spectra recorded at 4 K of **1** (black), **2** (green) and **3** (red). The powder simulation lines are plotted in blue with the different ZFS parameters for **1** ($D$ = 0.28 cm$^{-1}$ and $E$ = 0.018 cm$^{-1}$), **2** ($D$ = 0.29 cm$^{-1}$ and $E$ = 0.0025 cm$^{-1}$) and **3** ($D$ = 0.13 cm$^{-1}$ and $E$ = 0.013 cm$^{-1}$; a higher-order term $B_4^0 \hat{O}_4^0$ is also included in **Equation 1**, where $\hat{O}_4^0$ is the extended Stevens operator and $B_4^0$ = -1.5×10$^{-5}$ cm$^{-1}$). An isotropic $g$ factor of $g_{iso}$ = 1.99 and a ZFS strain ($Str_D$ = 0.006 cm$^{-1}$ and $Str_E$ = 0.003 cm$^{-1}$) accounting for inhomogeneous broadening are applied for all molecules. **b)** Echo-detected field-sweep spectra of **1**, **2** and **3** at 3.3 K. The blue lines represent the simulations of absorption spectra based on the same parameters extracted from CW fittings. The inset symbols indicate the transitions at different magnetic fields, which are further investigated for the spin dynamics. **c)** Zeeman splitting for the $S$ = 7/2 spin of **1**, **2** and **3** when the magnetic field is parallel to $z$-axis of the ZFS tensor (the splitting of $x$- and $y$-directions are illustrated in **Figure S3.2**). The blue circles show the positions of the transitions with 9.75 GHz microwave photons. The insets illustrate their binding sites.

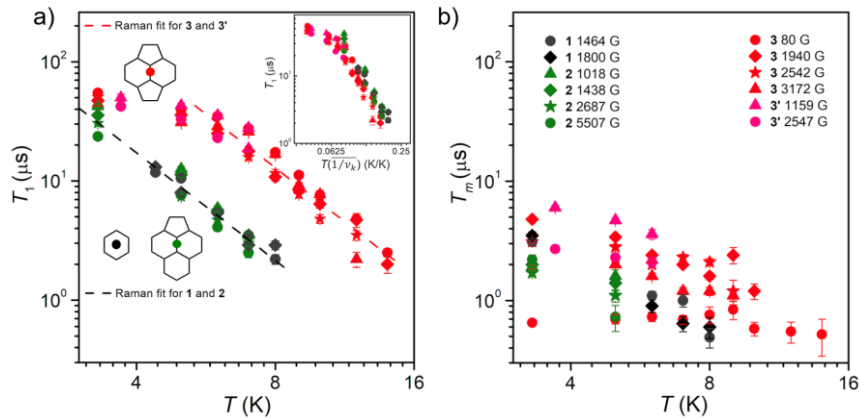

**Figure 3. a)** Temperature dependences of $T_1$ and **b)** $T_m$ of **1**, **2**, **3** and **3'** in $CS_2$ solution, showing with the same range of relaxation time and temperature. The inset diagrams display the three binding sites of pyracylene (**3** and **3'**, red), acephenalene (**2**, green) and hexagon (**1**, black). The symbols correspond to the peaks in EDFS spectra at different magnetic fields in **Figure 2**. The dashed lines in **a)** are Raman-type fits (see text). The inset figure in **a)** shows the frequency-normalized $T_1$-temperature dependence, which is achieved through multiplying temperature by the averaged reciprocal frequencies ($\overline{1/\nu_k}$, $k = 1$-3) of the first three vibrational modes. The unit of frequency is converted to K to obtain a dimensionless normalized temperature (K/K).

As far as $T_m$ is concerned, we observe that molecules **3** and **3'**, with rigid pyracylene binding sites, exhibit larger $T_m$ values than **1**, **2** (**Figure 3b**) and a larger dispersion among different magnetic fields than for $T_1$. No obvious conclusion can be extracted from this observation, other than the fact that $T_1$ and $T_m$ converge at higher temperatures, as is common in high-spin systems[18, 23]. This suggests that instead of nuclear spin diffusion, $T_m$ is limited by $T_1$ and consequently, the spin echo signal of the studied samples disappears above 15 K. In terms of actual numbers, below 4 K, all measurements result in $T_m > 1$ $\mu s$ except for **3** at 80 G, where the magnetic field is not strong enough to suppress spin-spin and spin-nuclei interactions[24]. The longest $T_m$ of 6.0(3) $\mu s$ is observed for **3'** at 1159 G and 3.7 K. The $T_m$ values are comparable with the long coherence times of spin-1/2 systems such as VOPc[25] and Y(Cp′)$_3$[8g], or of high-spin complexes based on $Cr^{3+}$ ($S = 3/2$)[23] and $Gd^{3+}$ ($S = 7/2$)[19]. The analysis is less straightforward, and the relaxation times are lower, when frozen $d^8$-toluene is employed as solvent instead of nuclear spin-free $CS_2$ (**Figure S3.15**). Apart from the presence of nuclear spins, toluene is also a larger and more anisotropically shaped solvent molecule, which is likely to present preferential solvation configurations to accommodate the shape of the EMFs. In practice, the methyl groups can be expected to present different typical distances to $Eu^{2+}$ in different EMFs, affecting $T_m$ in ways that are not directly related to the Eu binding site. Finally, the observation of Rabi oscillations in **3** (**Figure S3.15**) confirms that coherent manipulation of the spin is feasible to generate an arbitrary superposition of states within its ground spin multiplet.

**Spin-vibration coupling (SVC).** Motivated by the intriguing empirical relation between vibrational frequencies and the thermal behavior of $T_1$ depicted in the inset of **Figure 3a**, we performed further theoretical investigations of the first three metal-based vibrational modes and their relation to the spin Hamiltonian detailed in **Equation 1** (see **SI 4**). The goal here would be to offer an alternative explanation, independent of the vibrational energies and instead based on different SVC of different structures.

We started by combining well-established procedures to estimate SVC for different vibrational modes[1, 8d, i]: we carried out CASSCF *ab initio* calculations for **1**, **2** and **3** at progressively larger distortions $Q$ along the first three vibrational modes, which correspond to $x$, $y$, $z$ displacement of the Eu$^{2+}$ ion (see **Figure 4a**), since the other modes do not involve relevant changes in the Eu$^{2+}$-cage interaction and are higher in energy. Fitting the calculated spin sublevel energies to the ZFS Hamiltonian produces a plot of the dependence of $D$, $E$ with respect to the different vibrational distortions $Q_{1-3}$, leading to a qualitative conclusion that SVC of **3** seems to be overall reduced (**Figure 4b**). Based on these ZFS dependences on $Q_{1-3}$, the overall SVC strength for each mode can be also defined[26] (see **SI 4-2**) and calculated as shown in **Table S7**. This result further manifests the weaker SVC in **3**, which points in the same direction as the vibrational frequency results. Moreover, the coupling between vibrational modes and the idealized acoustic phonons in frozen solution is further probed (see **SI 4-3**) to show that the first three metal-based vibrations not only couple the most to the spin sublevels, but also can effectively facilitate the energy dissipation towards the thermal bath via the lattice phonons, making them the dominant factors that govern the spin relaxation process in this system.

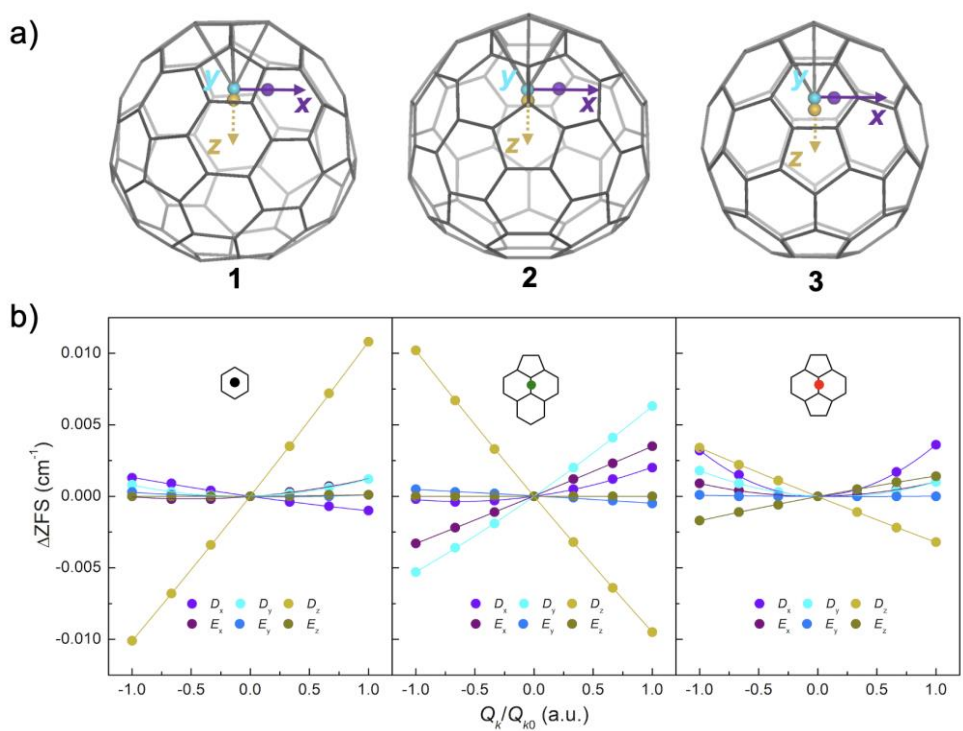

**Figure 4. a)** Illustration of metal distortions in **1** (left), **2** (middle) and **3** (right) along with vibrational mode 1 (*x*-direction, purple) and 3 (*z*-direction, yellow) from the equilibrium position (blue, the mode 2 is along with *y*-direction, perpendicular to the page). For clarity, the distorted metal sites indicated here are ten times of the distances of zero-point energies ($Q_{k0}$, $k = 1$ and 3) according to harmonic-oscillator approximation. **b)** The changes of ZFS parameters as a function of displacement for the three metal-based vibrational modes for **1** (left), **2** (middle) and **3** (right). The insets illustrate their binding sites. The solid lines in **b)** represent a guide for the eyes. The distortions $Q_k$ are normalized by $Q_{k0}$.

## CONCLUSION AND OUTLOOK

We have demonstrated for the first time the Eu(II)-based complex as a potential spin qubit candidate. More importantly, EMFs of this $4f^7$ lanthanoid in $CS_2$ solution have been used as ideal and simplest model systems to understand the spin dynamics of $Eu^{2+}$ inside these closed-shell carbon cages. In fact, some correlation between the structure and the spin dynamics can be established in a series of EMFs (**1**, **2**, **3** and **3'**) in which the binding sites on $Eu^{2+}$ are varying. One observes that **3** and **3'** with curved pyracylene binding site exhibit more rigid metal-based vibrations and slower spin relaxations than for **1** and **2**. Owing to the structural features in this family, vibrational frequency normalization of $T_1$ is feasible to produce a surprisingly good overlap between different compounds for $T_1$ values extending over an order of magnitude. This experimental observation evidences the importance of the low energy metal-based vibrations and how the energies track with temperature dependent $T_1$ data, a concept which should be general for any spin-bath environment and that supports previous theoretical works[8a, b]. With this general insight, future advances could aim at the fused-pentagon structure [12, 27] to hinder the metal-based vibrations. Given the protected high-spin state, the rationalized structure-relaxation correlation, the easy processability and the strong luminescence of $Eu@C_{2n}$ molecules, these may be further integrated into sophisticated qubit systems with their spin states coupled to the superconducting resonators[28], manipulated by an external electric field[29], or read out using light, in the same way it has been recently realized in $S$ = 1 Cr(IV)-based complexes[30].

## EXPERIMENTAL METHODS

**Synthesis and isolation.** $Eu@C_{2n}$ samples were produced using a modified arc-discharge method. Briefly, the anode graphite rod filled with $Eu_2O_3$/graphite powder with atomic ratio=1:20 was evaporated at 90 A under 300 torr helium static atmosphere. The soot was then refluxed in ortho-dichlorobenzene (*o*-DCB) under nitrogen atmosphere for 5 h, and followed by multi-step HPLC separation to yield pure samples. Positive-ion mode LD-TOF mass spectroscopy was used to check the sample purity. The detailed process is illustrated in Supporting Information.

**Spectroscopic characterizations.** Vis-near-infrared (Vis-NIR) absorption spectra were measured at room temperature in toluene solution on a Shimadzu 3100 spectrophotometer. The structures of obtained $Eu@C_{2n}$ samples were assigned by their absorption spectra compared with those of reported $M^{II}@C_{2n}$ (M = Eu and Sm) molecules (see **Figure S1.4** and **Table S1**). The photoluminescence (PL) spectra were collected at room temperature in $CS_2$ solution using a 375 nm laser excitation source and a spectrometer (Hamamatsu C9920-02 with a Hamamatsu PMA-11 optical detector). PL decays were measured using a compact fluorescence lifetime spectrometer (Quantaurus Tau C11367) and were fitted with a biexponential function. The excitation source was 365 nm LED. The PL lifetime measurement software U11487 was used to register the data.

**DFT calculations.** Geometry optimizations without symmetry restrictions of $Eu@C_{2n}$ molecules were carried out by Gaussian 09 package[31] using unrestricted hybrid density functional B3LYP with 6-31G(d) basis set for C atoms and Stuttgart−Dresden basis sets with effective core potential for Eu atoms, where 28 core electrons are included. Different metal sites, according to XRD structures, were tested in each case to achieve the most energetically favorable geometry. All optimized structures were proven by vibrational analyses to verify that the true local minimum is reached. Bonding analyses and the delocalization indices of $Eu@C_{2n}$ were obtained using the

Multiwfn program[32].

**EPR measurements.** The Eu@$C_{2n}$ samples were dissolved in deuterated $CS_2$ and $d^8$-toluene for EPR experiments. CW spectra were measured on a Bruker Elexsys E580 spectrometer operating in the X-band ($\omega$ = 9.47 GHz), whose spectra are simulated by EasySpin toolbox[33] (http://www.easyspin.org/) based on Matlab. Pulsed EPR data were collected on the same system by a 9.70 GHz cavity. The low-temperature environment was achieved by Oxford Instruments CF935 and ITC503 temperature controller. The signal of the pulsed-EPR experiments was collected by integrating the Hahn echo ($\pi/2$-$\tau$-$\pi$-$\tau$-echo). The $T_1$ values were measured by the inversion recovery method ($\pi$-$T$-$\pi/2$-$\tau$-$\pi$-$\tau$-echo) with 4-step phase cycling. The $T_m$ values were obtained by increasing the $\tau$ value of Hahn echo sequence with 2-step phase cycling. The $\pi/2$ and $\pi$ pulse lengths in EDFS, $T_1$ and $T_m$ measurements were 16 and 32 ns, respectively, with 10 dB attenuation of the microwave power. The Rabi oscillation experiments were carried out with a standard sequence ($t_p$−$T$−$\pi/2$−$\tau$−$\pi$−$\tau$−echo), where $T$ > 5$T_m$, by 8, 11, and 14 dB attenuation.

*Ab initio* **calculations.** CASSCF calculations were performed within OpenMOLCAS program package[34]. Scalar relativistic effects were considered with Douglas–Kroll–Hess transformation using ANO-RCC-VDZP basis set for all atoms. For the $f^7$ ground state of molecules at optimized structure and at each distorted geometry, generated from the three metal-based vibrational modes, the active space consists of seven electrons on the seven f-orbitals of $Eu^{2+}$. The molecular orbitals were optimized at the CASSCF level in a state-average (SA) over 188 doublet, 212 quartet, 48 sextet and an octet. The wave functions obtained at CASSCF level were then mixed by spin-orbit coupling (SOC) on all SA components by means of RASSI approach.

## ASSOCIATED CONTENT

### Supporting Information
HPLC profiles for the separation process, spectroscopic characterizations, CW- and pulsed EPR results of Eu@$C_{74-90}$ and theoretical details (PDF)

## AUTHOR INFORMATION

### Corresponding Authors
*zjshi@pku.edu.cn
*alejandro.gaita@uv.es
*eugenio.coronado@uv.es

### Notes
The authors declare no competing financial interest.

## ACKNOWLEDGEMENTS

This work is supported by the European Commission (ERC-2014-CoG-647301 DECRESIM, ERC-2018-AdG-788222 MOL-2D, the FET-OPEN project FATMOLS (No 862893) and the QuantERA project SUMO; the Spanish MICINN (grant CTQ2017-89993 co-financed by FEDER, grant MAT2017-89528; the Unit of excellence "María de Maeztu" CEX2019-000919-M); the Generalitat Valenciana (Prometeo Program of Excellence and PO FEDER Program IDIFEDER/2018/061, IDIFEDER/2020/063); the National Natural Science Foundation of China (grant No. 21875002); and the National Basic Research Program of China (grant No. 2017YFA0204901). We thank José

M. Martínez-Agudo, Gloria Agustí and Ángel López for their help with the EPR measurements, and Dr. Jesús Cerdá for helpful discussion and support in *ab initio* calculations.

**REFERENCE**

[1] F. S. Guo, B. M. Day, Y. C. Chen, M. L. Tong, A. Mansikkamaki and R. A. Layfield, *Science* **2018**, *362*, 1400-1403.

[2] A. Gaita-Arino, F. Luis, S. Hill and E. Coronado, *Nat. Chem.* **2019**, *11*, 301-309.

[3] a) M. Atzori and R. Sessoli, *J. Am. Chem. Soc.* **2019**, *141*, 11339-11352; b) M. R. Wasielewski, M. D. Forbes, N. L. Frank, K. Kowalski, G. D. Scholes, J. Yuen-Zhou, M. A. Baldo, D. E. Freedman, R. H. Goldsmith and T. Goodson, *Nat. Rev. Chem.* **2020**, 1-15.

[4] P. Stamp, *Philos. T. R. Soc. A* **2012**, *370*, 4429-4453.

[5] a) J. M. Zadrozny, J. Niklas, O. G. Poluektov and D. E. Freedman, *ACS. Cent. Sci.* **2015**, *1*, 488-492; b) C. J. Yu, M. J. Graham, J. M. Zadrozny, J. Niklas, M. D. Krzyaniak, M. R. Wasielewski, O. G. Poluektov and D. E. Freedman, *J. Am. Chem. Soc.* **2016**, *138*, 14678-14685.

[6] F. Moro, D. Kaminski, F. Tuna, G. F. Whitehead, G. A. Timco, D. Collison, R. E. Winpenny, A. Ardavan and E. J. McInnes, *Chem. Commun.* **2014**, *50*, 91-93.

[7] M. Shiddiq, D. Komijani, Y. Duan, A. Gaita-Arino, E. Coronado and S. Hill, *Nature* **2016**, *531*, 348-351.

[8] a) L. Escalera-Moreno, J. J. Baldovi, A. Gaita-Arino and E. Coronado, *Chem. Sci.* **2018**, *9*, 3265-3275; b) A. Lunghi, F. Totti, R. Sessoli and S. Sanvito, *Nat. Commun.* **2017**, *8*, 14620; c) M. Atzori, E. Morra, L. Tesi, A. Albino, M. Chiesa, L. Sorace and R. Sessoli, *J. Am. Chem. Soc.* **2016**, *138*, 11234-11244; d) L. Escalera-Moreno, N. Suaud, A. Gaita-Arino and E. Coronado, *J. Phys. Chem. Lett.* **2017**, *8*, 1695-1700; e) M. S. Fataftah, M. D. Krzyaniak, B. Vlaisavljevich, M. R. Wasielewski, J. M. Zadrozny and D. E. Freedman, *Chem. Sci.* **2019**, *10*, 6707-6714; f) R. Mirzoyan and R. G. Hadt, *Phys. Chem. Chem. Phys.* **2020**, *22*, 11249-11265; g) A. M. Ariciu, D. H. Woen, D. N. Huh, L. E. Nodaraki, A. K. Kostopoulos, C. A. P. Goodwin, N. F. Chilton, E. J. L. McInnes, R. E. P. Winpenny, W. J. Evans and F. Tuna, *Nat. Commun.* **2019**, *10*, 3330; h) A. Ullah, J. Cerdá, J. J. Baldoví, S. A. Varganov, J. Aragó and A. Gaita-Ariño, *J. Phys. Chem. Lett.* **2019**, *10*, 7678-7683; i) Y. Rechkemmer, F. D. Breitgoff, M. van der Meer, M. Atanasov, M. Hakl, M. Orlita, P. Neugebauer, F. Neese, B. Sarkar and J. van Slageren, *Nat. Commun.* **2016**, *7*, 10467; j) Y. Duan, J. T. Coutinho, L. E. Rosaleny, S. Cardona-Serra, J. J. Baldoví and A. Gaita-Ariño, *arXiv preprint arXiv:2103.03199* **2021**.

[9] a) J. J. L. Morton, A. M. Tyryshkin, A. Ardavan, S. C. Benjamin, K. Porfyrakis, S. A. Lyon and G. A. D. Briggs, *Nat. Phys.* **2005**, *2*, 40-43; b) R. M. Brown, Y. Ito, J. H. Warner, A. Ardavan, H. Shinohara, G. A. D. Briggs and J. J. L. Morton, *Phys. Rev. B* **2010**, *82*; c) Y.-X. Wang, Z. Liu, Y.-H. Fang, S. Zhou, S.-D. Jiang and S. Gao, *npj Quantum Inf.* **2021**, *7*, 1-8.

[10] A. A. Popov, S. Yang and L. Dunsch, *Chem. Rev.* **2013**, *113*, 5989-6113.

[11] S. Lebedkin, B. Renker, R. Heid, H. Schober and H. Rietschel, *Appl. Phys. A* **1998**, *66*, 273-280.

[12] L. Bao, Y. Li, P. Yu, W. Shen, P. Jin and X. Lu, *Angew. Chem. Int. Ed.* **2020**, *59*, 5259-5262.

[13] a) L. Bao, P. Yu, Y. Li, C. Pan, W. Shen, P. Jin, S. Liang and X. Lu, *Chem. Sci.* **2019**, *10*, 4945-4950; b) H. Yang, H. Jin, H. Zhen, Z. Wang, Z. Liu, C. M. Beavers, B. Q. Mercado, M. M. Olmstead and A. L. Balch, *J. Am. Chem. Soc.* **2011**, *133*, 6299-6306; c) M. Suzuki, Z. Slanina, N. Mizorogi, X. Lu, S. Nagase, M. M. Olmstead, A. L. Balch and T. Akasaka, *J. Am. Chem. Soc.* **2012**, *134*, 18772-18778.

[14] A. A. Popov and L. Dunsch, *Chem. – Eur. J.* **2009**, *15*, 9707-9729.

[15] R. C. Haddon, *J. Am. Chem. Soc.* **1987**, *109*, 1676-1685.

[16] A. Rodriguez-Fortea, N. Alegret, A. L. Balch and J. M. Poblet, *Nat. Chem.* **2010**, *2*, 955-961.

[17] a) H. Matsuoka, N. Ozawa, T. Kodama, H. Nishikawa, I. Ikemoto, K. Kikuchi, K. Furukawa, K. Sato, D. Shiomi and T. Takui, *J. Phys. Chem. B* **2004**, *108*, 13972-13976; b) M. D. Jenkins, Y. Duan, B. Diosdado, J. J. García-Ripoll, A. Gaita-Ariño, C. Giménez-Saiz, P. J. Alonso, E. Coronado and F. Luis, *Phys. Rev. B* **2017**, *95*.

[18] Z. Hu, B. W. Dong, Z. Liu, J. J. Liu, J. Su, C. Yu, J. Xiong, D. E. Shi, Y. Wang, B. W. Wang, A. Ardavan, Z. Shi, S. D. Jiang and S. Gao, *J. Am. Chem. Soc.* **2018**, *140*, 1123-1130.

[19] Z. Liu, H. Huang, Y.-X. Wang, B.-W. Dong, B.-Y. Sun, S.-D. Jiang and S. Gao, *Chem. Sci.* **2020**, *11*, 10737-10743.

[20] K. S. Pedersen, A. M. Ariciu, S. McAdams, H. Weihe, J. Bendix, F. Tuna and S. Piligkos, *J. Am. Chem. Soc.* **2016**, *138*, 5801-5804.

[21] M. Atzori, L. Tesi, S. Benci, A. Lunghi, R. Righini, A. Taschin, R. Torre, L. Sorace and R. Sessoli, *J. Am. Chem. Soc.* **2017**, *139*, 4338-4341.

[22] M. Atzori, S. Benci, E. Morra, L. Tesi, M. Chiesa, R. Torre, L. Sorace and R. Sessoli, *Inorg. Chem.* **2018**, *57*, 731-740.

[23] S. Lenz, H. Bamberger, P. P. Hallmen, Y. Thiebes, S. Otto, K. Heinze and J. van Slageren, *Phys. Chem. Chem. Phys.* **2019**, *21*, 6976-6983.

[24] M. K. Wojnar, D. W. Laorenza, R. D. Schaller and D. E. Freedman, *J. Am. Chem. Soc.* **2020**, *142*, 14826-14830.

[25] M. Atzori, L. Tesi, E. Morra, M. Chiesa, L. Sorace and R. Sessoli, *J. Am. Chem. Soc.* **2016**, *138*, 2154-2157.

[26] N. C. Chang, J. B. Gruber, R. P. Leavitt and C. A. Morrison, *The Journal of Chemical Physics* **1982**, *76*, 3877-3889.

[27] Y. Hao, L. Feng, W. Xu, Z. Gu, Z. Hu, Z. Shi, Z. Slanina and F. Uhlik, *Inorg. Chem.* **2015**, *54*, 4243-4248.

[28] M. D. Jenkins, D. Zueco, O. Roubeau, G. Aromi, J. Majer and F. Luis, *Dalton. Trans.* **2016**, *45*, 16682-16693.

[29] J. Liu, J. Mrozek, Y. Duan, A. Ullah, J. J. Baldoví, E. Coronado, A. Gaita-Ariño and A. Ardavan, *arXiv preprint arXiv:2005.01029* **2020**.

[30] S. L. Bayliss, D. W. Laorenza, P. J. Mintun, B. D. Kovos, D. E. Freedman and D. D. Awschalom, *Science* **2020**, *370*, 1309-1312.

[31] M. Frisch, G. Trucks, H. Schlegel, G. Scuseria, M. Robb, J. Cheeseman, G. Scalmani, V. Barone, B. Mennucci, G. Petersson, H. Nakatsuji, M. Caricato, X. Li, H.

P. Hratchian, A. F. Izmaylov, J. Bloino, G. Zheng, J. L. Sonnenberg, M. Hada, M. Ehara, K. Toyota, R. Fukuda, J. Hasegawa, M. Ishida, T. Nakajima, Y. Honda, O. Kitao, H. Nakai, T. Vreven, J. Montgomery, J. A., J. E. Peralta, F. Ogliaro, M. Bearpark, J. J. Heyd, E. Brothers, K. N. Kudin, V. N. Staroverov, R. Kobayashi, J. Normand, K. Raghavachari, A. Rendell, J. C. Burant, S. S. Iyengar, J. Tomasi, M. Cossi, N. Rega, N. J. Millam, M. Klene, J. E. Knox, J. B. Cross, V. Bakken, C. Adamo, J. Jaramillo, R. Gomperts, R. E. Stratmann, O. Yazyev, A. J. Austin, R. Cammi, C. Pomelli, J. W. Ochterski, R. L. Martin, K. Morokuma, V. G. Zakrzewski, G. A. Voth, P. Salvador, J. J. Dannenberg, S. Dapprich, A. D. Daniels, Ö. Farkas, J. B. Foresman, J. V. Ortiz, J. Cioslowski and D. J. Fox, *Gaussian 09, Revision A. 1* **2009**, Gaussian: Wallingford, CT.

[32] T. Lu and F. W. Chen, *J. Comput. Chem.* **2012**, *33*, 580-592.

[33] S. Stoll and A. Schweiger, *J. magn. reson.* **2006**, *178*, 42-55.

[34] I. Fdez. Galván, M. Vacher, A. Alavi, C. Angeli, F. Aquilante, J. Autschbach, J. J. Bao, S. I. Bokarev, N. A. Bogdanov and R. K. Carlson, *J. Chem. Theory Comput.* **2019**, *15*, 5925-5964.

**For Table of Contents Only**

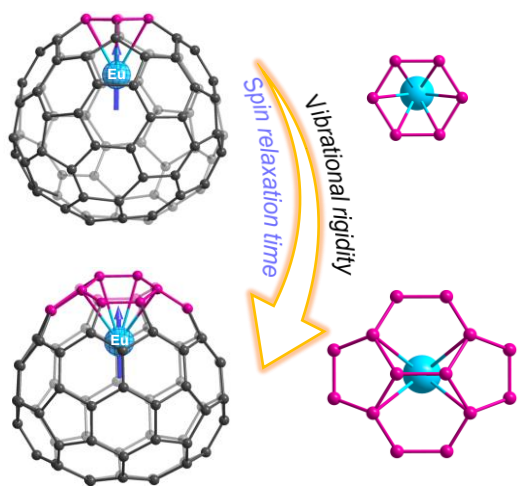